\documentclass[a4paper,12pt]{article}
\usepackage{latexsym}
\usepackage{amsmath}
\usepackage{amsfonts}
\usepackage{amssymb}
\usepackage{graphicx}
\usepackage{color}
\begin{document}
\title{Spatio-Temporal Patterns for a Generalized  Innovation Diffusion Model}

\author{{\,}\\{\large Fariba Hashemi} \\ {\scriptsize {\texttt Ecole Polytechnique F\'ed\'erale de Lausanne (EPFL), College of Management of Technology (CDM),}}\\ {\scriptsize {\texttt Management of Technology and Entrepreneurship Institute (MTEI), Station 5, CH-1015 Lausanne.}}\\ $\,$\\ Max-Olivier Hongler\\ {\scriptsize {\texttt Ecole Polytechnique F\'ed\'erale de Lausanne (EPFL), School of Engineering (STI),}}\\ {\scriptsize {\texttt Laboratory of Microengineering for Manufacturing (LPM), Station 17, CH-1015 Lausanne.}}\\ $\,$\\ Olivier Gallay\\ {\scriptsize {\texttt IBM Zurich Research Laboratory, Mathematical and Computational Sciences Department,}}\\ {\scriptsize {\texttt  Saeumerstrasse 4, CH-8803 Rueschlikon.}}\\{\,}\\}
\maketitle
\abstract{$\,$

\noindent We construct  a model of innovation diffusion that incorporates a spatial component into a classical imitation-innovation dynamics first introduced by F. Bass. Relevant for situations where the imitation process explicitly depends on the spatial proximity between  agents, the resulting nonlinear field dynamics is  exactly solvable. As expected for nonlinear collective dynamics,  the imitation mechanism generates spatio-temporal patterns, possessing here the remarkable feature that they can be  explicitly and analytically discussed. The simplicity of the model, its intimate connection with the original Bass' modeling framework  and the exact transient solutions  offer a rather unique theoretical stylized framework to describe how innovation jointly develops in space and time.}

\vspace{0.5cm}


\noindent 

\section{Introduction}
Since the middle of the XX$^{{\rm th}}$ century, a  substantial literature emphasizes the importance of quantitative models that enable the forecasting of the diffusion of technological innovations (DTI) not only for pure academic interest, but for its practical relevance. While the empirical evidence that quantification of DTI is possible was initially recognized by E. Mansfield \cite{mansfield} and Z. V.  Griliches \cite{griliches1, griliches2}, the first quantitative stylized dynamical modeling framework was  proposed by F. Bass in his seminal 1969 paper \cite{bass}. Similar in essence to  the P.-F.  Verhulst's  epidemiological  logistic equation, the  Bass' model uses  an {\bf aggregated differential approach} enabling the reproduction of the relevant dynamical features of the adoption of a new  product and/or a new technology in a society of consumers. Bass' nonlinear dynamics is basically governed by the ratio of two control parameters, namely the {\it innovation} and {\it imitation} rates.  Introducing a quadratic nonlinearity into the evolution equation, the model mathematically describes consumer imitative  interactions.  This nonlinearity leads to an evolution characterized by two distinct time scales, {\it i.e.} a fast initial exponentially growing phase followed by a slow asymptotic evolution when full equilibrium demand is nearly reached.  The seminal works of J. A. Schumpeter \cite{schumpeter} and  subsequently of B. Jovanovic and R. Rob \cite{jovanovic} support this  observation by illustrating the importance of imitation waves in DTI and the formation of  business cycles.\\
\\
\noindent In general, interaction-based models have been employed for various applications ranging from epidemiology  \cite{pastor},  variance of crime rates across space when direct interdependencies occur between nearest neighbors \cite{glaeser}, herd behavior in financial markets \cite{cont, corcos, daipra, horst1} to social movements and  political uprisings (see \cite{lopez-pintado1} and \cite{lopez-pintado2} for several additional  references).  Interaction-based methods have also been useful tools in modeling the diffusion of innovation \cite{valente, brock1, blume1, schultz, blume2, brock2}. In their contribution, G. Ellison and D. Fudenberg  \cite{ellison1, ellison-f}  study agents who consider the  experiences of their neighbors in deciding which of two technologies  to adopt,  in a world where players use Ôrules of thumbÕ that ignore  historical data but may incorporate a tendency to use the more popular  technology.  In this learning environment, agents observe both their neighbors' choices, and periodically reevaluate their decisions, as opposed to making a once-and-for-all choice. R.  Andergassen et al. \cite{andergassen}  investigate the evolutionary process of imitation and innovation as a search mechanism in a given neighbourhood of firms.  In their world, the spreading of information through neighbourhoods allows firms to acquire knowledge leading to innovation waves as firms attempt to glean information on best practice techniques, which they subsequently imitate.   For additional models where preference orderings over alternatives in a choice set can depend on the actions chosen by other agents, see \cite{levine, levinef, pesen,horst}.\\
\\
 \noindent  The original Bass' dynamics is aggregated and hence fails {\bf  to detail the influence of spatial  location of agents} in the imitative behavior of consumers.  Intuitively, spatial considerations strongly affect the interactions between agents and the underlying imitation mechanisms.    In fact spatial proximity is argued to be a major  driving force of innovation diffusion which  often exceeds  the  external marketing efforts such as advertising \cite{foster,rogers}. In 1991, P.  Krugman \cite{krugman} pointed out that production is remarkably concentrated in space. This observation opened a strong research effort  devoted to the understanding of the spatial dimension of  innovation diffusion. M. P. Feldman \cite{feldman1, feldman2} echoed P.  Krugman's  observation in pointing out that geographical effects are even more stringent  for innovative  activity because the rationale for the formation of adopter clusters is  related to the role of word-of-mouth and imitation in the diffusion of innovations.  As emphasized  in \cite{case}, a clear correlation  exists between geographical proximity and the strength and speed of word-of-mouth spread, sometimes labeled as the {\it neighborhood effect}.\\
\\
\noindent In addition to these works, a wealth of empirical  studies exemplify the importance of geography in the diffusion of  knowledge and  R \& D. Spatially-mediated knowledge spillovers of R \&  D are explicitly discussed in \cite{audretsch, acs1, acs2, jaffe}.   It is noteworthy to observe that this pure geographic view can be generalized by defining metric distances on abstract state spaces in order to  describe  the evolution of  technological advance, R \& D investment  volume or any other abstract features  \cite{akerlof} on which agents can compete by adjustment of their  individual behavior.\\

\noindent  Recent literature suggests that imitation interactions between interacting agents like bacteria,  flies, quadrupeds or fishes can explain the formation  of compact spatio-temporal patterns, {\it i.e.}  {\it swarms} or {\it platoons},   which spatially  evolve as quasi-solid bodies \cite{vicsek, cucker}. The {\it  flocking} mechanism  originates from mimetic type decisions  based on  agents' observations of their neighbors. To the best of our knowledge,
spatial flocking mechanisms seem  to be  barely discussed in the  interaction-based socioeconomic literature. Hence a natural and  simple attempt to analytically infer  the role of spatial parting   is to  introduce spatial effects and it is  the aim of our paper to incorporate their influence into  the original Bass' evolution. \\

\noindent Adding a spatial dimension transforms the Bass'  ordinary  differential equation into a partial differential equation (PDE).  Due to the underlying imitation mechanism, the resulting PDE will be  intrinsically nonlinear, a perspective that generally offers  little hope for explicit solutions in the realm of field theories.  The present paper illustrates how a simple natural spatial extension of the original Bass' dynamics  leads nevertheless to a fully solvable nonlinear field dynamics, a truly remarkable result. The resulting equations  belong  to the {discrete velocities Boltzmann equations} (DVBE) which describe the  macroscopic properties of a dilute gas.  The specific DVBE that can be  derived from the Bass' dynamics coincide with the Ruijgrok-Wu  (RW) model introduced and solved  by  T. W.  Ruijgrok and T. T. Wu \cite{ruijgrok}. This intimate connection with statistical physics suggests that the Bass' dynamics can be obtained, via a {\it  mean-field} limit, from a microscopic point of view in which a large  number of agents interact.   While the mean-field approach  is  a basic tool in statistical physics of  large systems, it has now been explicitly used in recent econometric studies as well (see illustrations in \cite{corcos,daipra, horst1,  lopez-pintado2, collet, horst}).  Relying on the  law of large numbers,  the mean-field limit allows one to write  deterministic evolution for probability densities in question.  In  the sequel, we will explicitly construct the microscopic connection that exists between the Bass' imitation model with spatial effects  and the RW model inspired by  a similar approach  adopted in \cite{hongler1} for a multi-agent dynamics  in logistics and econophysics contexts.  In \cite{hongler1}, the dynamics exhibit a nonlinear term due to a specific imitation mechanism giving  rise to the famous Burgers' nonlinear PDE to describe the emergence of spatio-temporal patterns. As illustrated in \cite{hongler2}, RW dynamics actually generalizes the Burgers' equation, the spatio-temporal Bass'  model presented here can itself be viewed as a natural generalization  of the multi-agent imitation model studied in \cite{hongler1}.  \\

\noindent  Besides its direct practical relevance, the simplicity of the original BassÕ model, for which exact analytical  solutions are available, has undoubtedly contributed to its  popularity in the economics and management literatures. Endowing Bass' dynamics with spatially-dependent imitation mechanisms confers a new  dimension to interaction-based socioeconomic modeling, opening the  possibility to analytically  study the generation of spatio-temporal  patterns in a highly nonlinear context.  Our stylized dynamics can be  viewed as an exceptional possibility to  analytically observe the spatio-temporal effects arising for a  collection of agents subject to imitation interactions. 
\section{Spatially-Dependent Imitation Dynamics}

Consider a collection ${\cal A}$ of $N$ autonomous agents  which are  in a migration process on the one-dimensional real line $\mathbb{R}$. At any time $t\in \mathbb{R}^{+}$, we assume that the complete population  is composed by two types of agents ${\cal A}_+$ and ${\cal A}_-$, (${\cal A}={\cal A}_-\cup {\cal A}_+$),  characterized  by two associated $(x,t)$-dependent migration velocities $V_+(x,t)$ and $V_-(x,t)$ on $\mathbb{R}$.  At any time, each agent is free to modify his/her velocity.  Agents $a_k\in {\cal A}$, $k=1,2,...,N$, flip their velocities  either spontaneously or after an autonomous decision based on an imitation process (IP).  Let us  write $\alpha(x,t)$, respectively $\beta(x,t)$,  as the spontaneous transformation rates  from states ${\cal A}_+ \mapsto_{\alpha(x,t)} {\cal A}_-$, respectively ${\cal A} _-\mapsto_{\beta(x,t)} {\cal A}_+$. Apart from these spontaneous transitions, additional  transitions are assumed to  be  triggered by mutual agents' interactions. Specifically,  for agent $a_k$, the  IP mechanism  is assumed to depend on the observation of the present velocity states,  ({\it i.e.} $V_+$ or $V_-$), adopted by other proximity members located in the neighborhood ${\cal N}_k = {\cal N}_{k,-}\cup {\cal N}_{k,+} \subset {\cal A}$ of agent $a_{k}$.   For an arbitrary agent  $a_{k ,\pm}$,  we define his/her imitation decision rule according to his/her interactions with other agents as follows:

\begin{itemize}
  \item[] {\it i)} {\bf Dynamic rule for an agent $\boldsymbol{a_{k,-}}$}.  At time $t$, the agent $a_{k,-}$ simultaneously observes the (velocity) states of the agents contained in his/her neighborhood ${\cal N}_k$.  The presence  of agents $a_{j,+}\in {\cal N}_k$, $j\neq k$, seen by $a_{k,-}$  triggers an imitation mechanism which  enhances the transition rate towards state $V_+$, {\it i.e.}  $\beta(x,t) \mapsto \left[\beta (x,t) + i_k(x,t)\right]$. The extra contribution  $i_{k}(x,t)$ is proportional to (monotonically increasing with) the number $N_{k}$ of agents $a_{j,+}$  in the neighborhood ${\cal N}_k$ of agent $a_{j,-}$ ({\it i.e.} $N_{k}=\mathrm{card}\left({\cal N}_{k,+}\right)$).
  \item[]
  \item[] {\it ii)} {\bf Dynamic rule for an agent $\boldsymbol{a_{k,+}}$}. At time $t$, the agent $a_{k,+}$ simultaneously  observes the (velocity) states of the agents contained in his/her neighborhood ${\cal N}_k$ and {\bf does not modify its velocity} whatever he/she observes.
\end{itemize}

\noindent According to these dynamic rules, the time-dependent position $X_k(t)$ of agent $a_{k}$, $k=1,2,...,N$,  can be written as a set of coupled stochastic differential equations (SDEs):

\begin{equation}
\label{INDIVDUAL}
\dot{X}_k(t) = I_k(t), \qquad k=1,2,...,N,
\end{equation}

\noindent where $I_k(t)$ stands for a two-velocity-states Markov chain ({\it i.e.} the state space is here $\Omega:= \left\{ V_-(x,t), V_+(x,t)\right\}$) the transition rates of which are defined by :
\begin{equation}
\label{TRANSOS}
 V_+(x,t) \mapsto_{\alpha(x,t)} V_-(x,t) \qquad V_- (x,t)\mapsto_{\left[ \beta(x,t) + i_k(x,t) \right]} V_+(x,t). 
\end{equation}

\noindent The noise source in Eq.(\ref{INDIVDUAL}) can also be viewed  as a non-homogeneous, alternating Markov renewal process in which the inverse transition rates $\alpha(x,t)^{-1}$ and $\left[\beta(x,t) + i_k(x,t)\right]^{-1}$ are respectively the mean sojourn times in states $V_{+}(x,t)$ and $V_{-}(x,t)$.  Observe therefore that  in Eq.(\ref{INDIVDUAL}), the coupling between the various agents is realized via the extra $i_k(x,t)$ transition rate.\\
\\
\noindent In what follows, we consider a very large population of agents, {\it i.e.} $N\rightarrow \infty$, and we focus on observation neighborhoods ${\cal N}_k$, $k=1,2,...,N$, that contain on average  $N_k=\rho N$ agents ($0< \rho <1$), implying in turn that $N_k \rightarrow \infty$. Instead of individual trajectories, let us now think in terms of statistical properties characterizing the evolution of  the whole population of agents ${\cal A}$. To this aim, we  write $P(x,t)\in [0,1]$ and $Q(x,t)\in[0,1]$  to denote the density of agents ${\cal A}_+$ and ${\cal A}_-$ to be found at position $x \in \mathbb{R}$ at time $t$. \\
\\
\noindent We now adopt a {\bf mean-field approach (MFA)}, which consists in considering that, statistically, the time evolution of an arbitrary agent, say $a_k \in {\cal A}$,  is representative of the whole population and the  influence due to  his/her neighbors located in ${\cal N}_k$ is viewed  as an {\bf external effective interactive mean field}.  Accordingly, the MFA allows us to replace Eq.(\ref{INDIVDUAL}) by  a single scalar SDE:

\begin{equation}
\label{SCALAR}
\dot{X}(t) = I(t), 
\end{equation}

\noindent where in Eq.(\ref{SCALAR}), the $i_k(x,t)$ transition rates are replaced by an effective rate $i(x,t)$ reading as:

\begin{equation}
\label{RATRI}
i(x,t) = \lim_{N \rightarrow \infty}{1 \over \rho N}\sum_{j=1}^{\rho N} a_{j, +}=\lim_{{\cal N}\rightarrow \infty}{1 \over {\cal N}}\sum_{j=1}^{{\cal N}} a_{j, +} \simeq \int_{x-\Gamma/2}^{x+\Gamma/2}  P(z,t) dz.
\end{equation}

\noindent where the radius  $\Gamma \in \mathbb{R}^{+}$ characterizes the size of the typical neighborhood interval ${\cal N}$ where imitations occur. \\
\\
\noindent Note that while we proceeded here heuristically, a rigorous mathematical justification for the validity of the MFA for this type of dynamics relies on  the property of  {\it propagation of chaos}, as  explained in \cite{gutkin} for SDEs of the type given by Eq.(\ref {INDIVDUAL}) when the noise is a White Gaussian Noise (WGN). SDEs of the type given by Eq.(\ref {INDIVDUAL}) driven by WGN have recently been used for multi-agent dynamics with imitation process in \cite{hongler1}.\\
\\
\noindent Associated with the SDEs given by Eqs.(\ref{SCALAR}) and (\ref{RATRI}),  the MFA enables us to write a Fokker-Planck equation for the probability densities $P(x,t)$ and $Q(x,t)$, \cite{hongler2}:

$$
\dot{P}(x,t) +  V_{+}(x,t) {\partial \over \partial x}P(x,t) = +{\cal J} \left(P(x,t), Q(x,t)\right)- \alpha P(x,t) + \beta Q(x,t), \quad \quad  
$$
\begin{equation}
\label{RATEMF}
\dot{Q}(x,t) + V_{-}(x,t) {\partial \over \partial x}Q(x,t) = - {\cal J} \left(P(x,t), Q(x,t)\right) + \alpha P(x,t) - \beta Q(x,t),
\end{equation}

\noindent where  the {\bf imitation rate term} in Eqs.(\ref{RATEMF}) reads as:
\begin{equation}
\label{COLLISEUM}
{\cal J} \left(P(x,t), Q(x,t)\right)=Q(x,t)\left[\int_{x-\Gamma/2}^{x+\Gamma/2}  P(z,t) dz \right].
\end{equation}

\vspace{0.2cm}
\noindent The dynamics in Eqs.(\ref{RATEMF}) is a coupled  set of {\bf nonlocal and nonlinear field equations}, which barely offers  hope for any analytical discussion. However, for small $\rho$ implying small radius $\Gamma$ ({\it i.e.} infinitesimal  interaction neighborhoods),  we may Taylor expand Eqs.(\ref{RATEMF}), up to first order in $\Gamma$,    to obtain:

$$
\dot{P}(x,t) +  V_{+}(x,t) {\partial \over \partial x}P(x,t) = + \Gamma P(x,t) Q(x,t)  - \alpha P(x,t) + \beta Q(x,t), \quad\,\,  
$$
\begin{equation}
\label{SPATIO-BASS}
\dot{Q}(x,t) + V_{-}(x,t) {\partial \over \partial x}Q(x,t) = - \Gamma P(x,t) Q(x,t) + \alpha P(x,t) - \beta Q(x,t).
\end{equation}

\noindent We directly observe that Eqs.(\ref{SPATIO-BASS}) can be viewed as  a generalized Bass' dynamics which confers relevance to  the spatial dimension on which agents evolve. Indeed, similarly to  the original Bass' model, we include an imitation process, represented in 
Eqs.(\ref{SPATIO-BASS}) by the nonlinear contribution $\Gamma P(x,t) Q(x,t)$. \\
\\
\noindent Writing $P(x,t) + Q(x,t) = \Sigma(x,t)$, the summation of both equations in (\ref{SPATIO-BASS}) yields a continuity equation:
\begin{equation}
\label{CONTINUITY}
{\partial \over \partial t} \Sigma(x,t)+  \left[V_{+}(x,t) + V_{-}(x,t) \right]{\partial \over \partial x}  \Sigma(x,t)=0, 
\end{equation}

\noindent which in turn implies the normalization constraint:

\begin{equation}
\label{NORMAL}
\int_\mathbb{R} P(x,t) dx + \int_\mathbb{R} Q(x,t) dx  \equiv 1,\quad \forall t \in \mathbb{R}^{+}.
\end{equation}

\noindent Although  the dynamics given by Eqs.(\ref{SPATIO-BASS}) has been derived for non-homogeneous and non-stationary parameters  $\alpha(x,t)$, $\beta(x,t)$ and $V_{\pm}(x,t)$, in the sequel we will restrict our attention  to situations where these parameters can  be assimilated to constants ({\it i.e.} $\alpha,\beta$ and   $V_{\pm}$).

\subsection{Spatial Homogeneous Regimes - Bass' Model}

\noindent  When $V_-= V_+ = 0$, the spatial character disappears from the dynamics given by Eqs.(\ref{SPATIO-BASS}), which implyies that $P(x,t) \equiv P(t)$ and $(Q(x,t) = Q(t)$. More precisely, Eqs.(\ref{SPATIO-BASS}) becomes:

$$ \dot{P} (t) = +\Gamma P(t) Q(t) - \alpha P(t) + \beta Q(t), $$
\begin{equation}
\label{BASSM}
 \dot{Q} (t) = -\Gamma P(t) Q(t) + \alpha P(t) - \beta Q(t), 
\end{equation}

\noindent with the notation $\dot{P}(t):= {d \over dt} P(t)$. The constraint given by Eq.(\ref{NORMAL}) now simply becomes $P(t)+Q(t)=1$ and enables us to rewrite Eq.(\ref{BASSM}) in the following form:

\begin{equation}
\label{BASSM1}
 \dot{P} (t) = -\Gamma \left\{ P^{2}(t) - \left[ 1 - {(\alpha+ \beta)\over \Gamma} \right]P(t) -{ \beta \over \Gamma}\right\}. 
\end{equation}

\noindent At this stage, it is worth  observing that for $\alpha=0$ and $\Gamma=1$,  Eq.(\ref{BASSM1})  reduces to:

\begin{equation}
\label{BASSM2}
 \dot{P}(t) = \left[1-P(t) \right]\, \left[\beta+  P(t) \right],
  \end{equation}
  
  \noindent which is precisely  the original Bass' dynamics \cite{bass} with $\beta$ being the ratio between the {\it imitation} and {\it innovation} rates.\\
  \\
  \noindent For completeness of the exposition, let us integrate Eq.(\ref{BASSM1}), with  the initial condition $P(t=0)= P_0$, to get:
  
  \begin{equation}
\label{SOLU_BASS}
P(t) = {( \Delta+b) \left( P_0 -\Delta+b \right) e^{- 2 \Delta \Gamma t}+ (\Delta-b) \left( P_0 +  \Delta+b  \right) \over \left(P_0 +  \Delta+b\right) - \left( P_0  - \Delta+b \right) e^{- 2 \Delta \Gamma t}}
\end{equation}

\noindent with the definitions:

$$
b=- {1 \over 2\Gamma}\left(\Gamma - \alpha - \beta \right)\quad{\rm and} \quad\Delta = {1 \over 2\Gamma}\sqrt{\left(\Gamma -\alpha-\beta \right)^{2} + 4 \beta\Gamma}.
$$

\noindent In the asymptotic time limit $t \rightarrow \infty$, Eq.(\ref{SOLU_BASS}) converges to:

\begin{equation}
\label{PSBASS}
\lim_{t\rightarrow \infty} P(t) =P_{{\rm station}}= \Delta-b.
\end{equation}

\noindent For the original Bass' model obtained when $\alpha=0$ and $\Gamma=1$, we have $b=- {1\over 2}(1-\beta)$ and $\Delta = {1\over 2}(1+\beta)$. Accordingly, with $P_0=0$, Eq.(\ref{SOLU_BASS}) reduces to the original Bass' solution:

\begin{equation}
\label{BASSORI}
P(t) = {\beta \left[ 1-  e^{-(1+\beta)t} \right] \over \beta + e^{-(1+\beta)t}}, \qquad Q(t) =  {(1+\beta)  e^{-(1+\beta)t}\over \beta + e^{-(1+\beta)t}}.
\end{equation}

\noindent Moreover $P_{{\rm station}}= \Delta- b=1$, which expresses the fact that all agents ultimately  adopt the new technology, as it is obviously expected in the original Bass' modeling framework, when  $\alpha=0$. 
\section{Bass' Dynamics with Spatio-Temporal Effects}

\noindent Coming back to the general dynamics given by Eqs.(\ref{SPATIO-BASS}) and introducing dimensionless coordinates via the Galileo transformation   $\left( x, t \right) \mapsto \left( y, s \right)$ defined by:
\begin{equation}
\label{GALLI}
\begin{pmatrix}
      y \\
      s  
\end{pmatrix} = \begin{pmatrix}
   {  2\Gamma \over (V_{+} -V_-)} &  {\Gamma (V_- + V_+)\over   (V_- -V_+)}\\
     0 &  \Gamma
\end{pmatrix}
\begin{pmatrix}
      x    \\
      t
\end{pmatrix},
\end{equation}

\noindent we straightforwardly have:

\begin{equation}
\label{DIFFOP}
\partial_x (\cdot) \mapsto \left[{2\Gamma\over (V_{+} - V_-)} \right] \partial_y  (\cdot) \quad {\rm and} \quad \partial_t  (\cdot)\mapsto \left[{\Gamma(V_-  +V_+)\over (V_- - V_+)} \right] \partial_y (\cdot) + \Gamma\partial_s (\cdot),
\end{equation}
\noindent  which  transforms Eqs.(\ref{SPATIO-BASS}) into the coupled set of nonlinear partial differential equations (PDEs):

$$
\partial_s P +\partial_y P =+ PQ - {\alpha\over \Gamma} P + {\beta\over \Gamma} Q, $$
\begin{equation}
\label{RW}
\partial_sQ -\partial_y Q = -PQ + {\alpha \over \Gamma}P - {\beta\over \Gamma} Q.
\end{equation}

\noindent The set of non-linear PDEs given by Eqs.(\ref{RW}) can be interpreted as being a discrete two velocity model of Boltzmann equations, first studied in \cite{ruijgrok}. The dynamics given by Eqs.(\ref{RW}) is remarkable, as using a generalized Hopf-Cole logarithmic transformation, we get:

 \begin{equation}
 \label{LOGOS1}
P(y,s) = - {\beta\over \Gamma} + \partial_s \log H(y,s)  - \partial_y \log H(y,s)
\end{equation}

 \begin{equation}
\label{LOGOS2}
 \quad Q(y,s)= {\alpha\over \Gamma} - \partial_s \log H(y,s)  - \partial_y \log H(y,s)\quad\,\,\,
\end{equation} 

\noindent which actually reduces the set Eqs.(\ref{RW})  into the {\it Telegraphist equation}:

\begin{equation}
\label{LOGOS3}
\partial_{ss} H(y,s)  - \partial_{yy} H(y,s) - {\alpha \beta \over \Gamma^{2}} H(y,s)=0.
\end{equation}

\noindent Accordingly,  the dynamics given by Eqs.(\ref{SPATIO-BASS}) with constant parameters has the {\bf l property that it can  be exactly solved for any initial conditions} $P_0(y)$ and $Q_0(y)$. According to \cite{ruijgrok}, the general solution reads as:
\begin{equation}
\label{BESSEL}
H(y,s) = {1\over 2}\left[ A(y+s) + A(y-s)\right] + {1\over 2} {\cal B}_1(y,s) + {\eta s\over 2} \,  {\cal B}_2(y,s),
\end{equation}

\noindent  where $\eta = {\sqrt{\alpha \beta}  \over  \, \Gamma}$ and where we have the following definitions:
$${\cal B}_1(y,s) = \int_{y-s}^{y+s} \mathbb{I}_0\left(\eta \sqrt{s^{2} - (y-y')^{2}} \right) B(y')dy',$$

\noindent
and 

$$
{\cal B}_2(y,s) = \int_{y-s}^{y+s}  \left({ 1 \over \sqrt{s^{2} - (y-y')^{2}}}\right) \mathbb{I}_1\left(\eta \sqrt{s^{2} - (y-y')^{2}} \right) A(y')dy',
$$

\noindent with $\mathbb{I}_0(\cdot)$ and $\mathbb{I}_1(\cdot)$ being the  modified Bessel's functions and:

\begin{equation}
\label{INIT1}
B(y) =  {1\over2} \left[P_0(y) - Q_0(y) + {\alpha \over \Gamma} + {\beta \over \Gamma} \right] A(y),
\end{equation}

\begin{equation}
\label{INIT2}
A(y) =  \exp\left\{ - {1\over2}\int_{0}^{y} \left[P_0(y') + Q_0(y') - {\alpha \over \Gamma} + {\beta \over \Gamma} \right]dy'\right\}.
\end{equation}

\subsection{Behavior of the Solutions}

\noindent Though fully explicit and exact, the solution given by Eqs.(\ref{LOGOS1}) and (\ref{LOGOS2}) deserves discussion and interpretation for specific situations and this is precisely the objective of this section.  In what follows, we will systematically choose $\Gamma=1$ and unit velocities $V_{\pm} = \pm1$.\\
\\
\noindent Let us here focus on the Bass' dynamics that results when $\alpha=0$. This directly  implies that  $\eta = \sqrt{\alpha\beta}=0$ and hence  Eqs.(\ref{LOGOS1})  and (\ref{LOGOS2}) coincide with the ordinary wave equation.   By definition of Bessel's functions \cite{abramovitz}, we have that $\mathbb{I}_0(0)=1$ and  $\mathbb{I}_1(0)=0$, thus  leading to the usual wave solution in the form:

\begin{equation}
\label{EQONDE}
H(y,s) = {1\over 2}\left[ A(y+s) + A(y-s)\right] + {1\over 2} \int_{y-s}^{y+s} B(y') dy'.
\end{equation}

\noindent Due to the fact that $\alpha =0$, we expect that the agents' population ${\cal A}_+$ with velocity $V_+=+1$ increases by opposition to the population ${\cal A}_-$ of agents with velocity $V_-=-1$ which is doomed to extinction. Let us now explore the transient nature of the solution and this  for  three types of initial conditions.

\begin{itemize}
  \item[] {\bf a)} {\bf  Initial conditions:  $\boldsymbol{P_0(y)} \boldsymbol{=} \boldsymbol{\delta(y)}$  and  $\boldsymbol{ Q_{0}(y)}\boldsymbol{=}\boldsymbol{0}$}.

  \noindent In  this case, the time-dependent solution reads as:
\begin{equation}
\label{CONS1}
P(y,s) = \delta(y-s) \qquad {\rm and } \qquad Q(y,s) \equiv 0.
\end{equation}  

\noindent as it can be checked directly from Eqs.(\ref{LOGOS1}) and (\ref{LOGOS2}).
Eqs.(\ref{CONS1}) are clearly consistent with the fact that  $\alpha =0$ and hence that no transitions from $V_+=+1$ to $V_-=-1$ velocities occur. Hence,  starting with all agents with velocity $V_+=+1$, they stay with their original velocity and the density $P(y,s)$  is a uniformly traveling Dirac mass with velocity $V_+=+1$ towards the positive  $\mathbb{R}$-axis.

\vspace{0.35cm}
  \item[] {\bf b)} {\bf Initial conditions:  $\boldsymbol{P_{0}(y)}\boldsymbol{=}\boldsymbol{0}$ and $\boldsymbol{Q_0(y)} \boldsymbol{=} \boldsymbol{\delta(y)}$}.

  \noindent Either by  direct substitution into Eqs.(\ref{RW}) or alternatively  by using  Eqs.(\ref{LOGOS1}) and (\ref{LOGOS2}), one can verify that the time-dependent solution in this case reads as:
  \begin{equation}
\label{CONS2}
P(y, s) = {\beta\over 2} e^{{\beta\over 2}(y-s)} \Theta(|y|-s)\quad {\rm and }\quad Q(y,s) = e^{-\beta s} \delta(y+s)
\end{equation}

\noindent where $\Theta(|y|-s)$ is a Heaviside cutoff function which identically vanishes for negative 
\begin{figure}[h]
\begin{center}
\includegraphics[width=14.4 cm,height=7.8 cm]{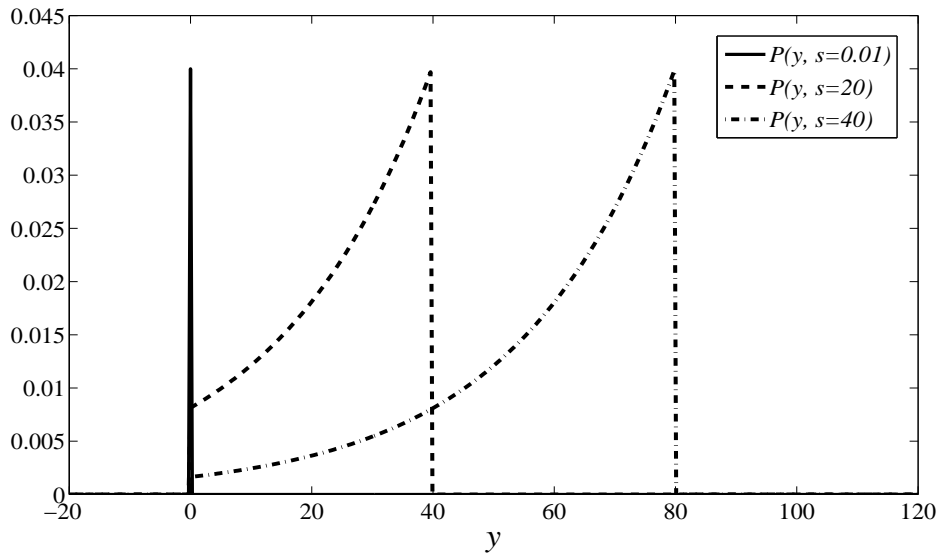}
\end{center}
\caption{Spatial dispersion $P(y, s)$ of the agents with velocity $V_+=+1$ for different times $s=[0.01; 20; 40]$. The initial conditions are given by $P_{0}(y)=0$ and $Q_{0}(y)=\delta(y)$, $\alpha=0$ and $\beta=0.1$. In this case, the agents are immediately segregated and there are no imitation processes.} \label{no_interaction}
\end{figure}
arguments. The dynamics $P(y, s)$ of the spatial dispersion of the agents with velocity $V_+=+1$ is illustrated in Fig. \ref{no_interaction}. \\
\\

\noindent   The behavior of the solution given by Eqs.(\ref{CONS2}) can be easily understood. Indeed, the Dirac mass for $Q(y,s)$ expresses the fact that agents with velocity $V_-=-1$ are gradually depopulated at the rate $s\beta $ for the benefit of agents traveling  with  velocity $V_+=+1$ and hence migrating to  $P(y,s)$. The Heaviside function $\Theta(|y|-s)$ expresses the fact that no agent can possibly be found at a distance larger than $|y|$ at time $t$ (remember that the velocities  here are $V_\pm=\pm1$).  It is worth  observing that in the original Bass' model, the adoption rate  given in Eq.(\ref{BASSORI}) is equal to $(1+\beta)$. This  overcomes the $\beta$ adoption rate that we will obtain below in Eqs.(\ref{LIMOSS}), due to the fact that in the present configuration, imitation process does not enter into play in  our spatial model, which drastically moderates the overall adoption rate. Indeed, in this limiting regime the agents' populations ${\cal A}_+$ and ${\cal A}_-$ are immediately and definitively segregated, which hence never allows imitation processes to take effect (the unit rate discrepancy between the $(1+ \beta)$ adoption rate occurring in  Eq.(\ref{BASSORI}) and the  $\beta$ rate  in Eq.(\ref{CONS2}) is obtained from the  $\Gamma=1$ choice). The resulting temporal evolution of the agents' overall adoption rate obtained in the present case ($P(t)= \int_{\mathbb{R}}  P(y,t) dy$), compared with the one of the original Bass' model, will be  illustrated later in Fig. \ref{bass_comparison}.\\
\\

\noindent Finally, let us observe that, from Eqs. (\ref{CONS2}), one immediately obtains:
\begin{equation}
\label{LIMOSS}
 \int_{\mathbb{R}}  P(y,s) dy = \left[ 1- e^{-\beta s}\right] \quad{\rm and }\quad  \int_{\mathbb{R}}  Q(y,s) dy =  e^{- \beta s},
\end{equation}

\noindent thus showing that Eq.(\ref{NORMAL}) is fulfilled.  In addition, for asymptotic times $s\rightarrow \infty$, Eqs.(\ref{LIMOSS})  indicate that all agents ultimately adopt the $V_+=+1$  velocity as it is expected for the $\alpha=0$ regime.

\vspace{0.35cm}
 \item[] {\bf c)} {\bf Initial conditions:}  $\boldsymbol{P_0(y)} \boldsymbol{=} \boldsymbol{Q_{0}(y)} \boldsymbol{=} \boldsymbol{\frac{1}{8\gamma}}\boldsymbol{[}\boldsymbol{\mathrm{tanh}}\boldsymbol{(y+\gamma)}\boldsymbol{-}\boldsymbol{\mathrm{tanh}}\boldsymbol{(y-\gamma)}\boldsymbol{]}$ {\bf .}\\
For these initial conditions and for the particular choice $\beta=2$ and $\gamma=\frac{1}{4}$, the resulting time-dependent solution of our spatial Bass' model is given by:
 $$P(y,s)= \frac{1}{2}\left\{-2+\frac{1}{H(y,s)}\left[A(y-s)\,\,+\right. \right.\qquad\qquad\qquad\qquad\qquad\qquad$$
\begin{equation}
\label{interaction_1}
\qquad\left.\mathrm{e}^{-y+s}\frac{\mathrm{cosh}(y-s-\gamma)}{\mathrm{cosh}(y-s+\gamma)}\left(1-\mathrm{tanh}(y-s-\gamma)+\mathrm{tanh}(y-s+\gamma)\right)]\right\}
\end{equation}
and
$$Q(y,s)= \frac{1}{2}\left\{\frac{1}{H(y,s)}\left[-A(y+s)\,\,+\right. \right.\qquad\qquad\qquad\qquad\qquad\qquad$$
\begin{equation}
\label{interaction_2}
\qquad\left.\mathrm{e}^{-y-s}\frac{\mathrm{cosh}(y+s-\gamma)}{\mathrm{cosh}(y+s+\gamma)}\left(1-\mathrm{tanh}(y+s-\gamma)+\mathrm{tanh}(y+s+\gamma)\right)]\right\},
\end{equation}
where 
$$H(y,s)=\frac{1}{2}\left[A(y+s)+A(y-s)+{\cal B}_1(y,s)\right], \qquad\qquad\qquad\qquad$$
$$A(y)= e^{y} \exp\left\{ \int_{0}^{y}\left[ \tanh\left( z- \gamma \right) - \tanh\left( z+ \gamma\right)\right] dz \right\}=
\mathrm{e}^{y}\frac{\mathrm{cosh}(y-\gamma)}{\mathrm{cosh}(y+\gamma)}, \qquad\qquad\qquad\qquad\qquad\qquad\qquad\qquad\qquad$$
$${\cal B}_1(y,s)=\frac{\mathrm{e}^{-\gamma}}{2}\left[\mathrm{atan}\left(\mathrm{sinh}(y+s+\gamma)\right)-\mathrm{atan}\left(\mathrm{sinh}(y-s+\gamma)\right)\right]\,\,$$
$$\qquad\qquad\quad-\mathrm{e}^{3\gamma}\left[\mathrm{e}^{-y-s-\gamma}+\mathrm{atan}(\mathrm{e}^{y+s+\gamma})-\mathrm{e}^{-y+s-\gamma}-\mathrm{atan}(\mathrm{e}^{y-s+\gamma})\right].$$
The spatio-temporal dynamics $P(y, s)$  of the agents with velocity $V_+=+1$ is illustrated in 
\begin{figure}[h]
\begin{center}
\includegraphics[width=14 cm,height=8.5 cm]{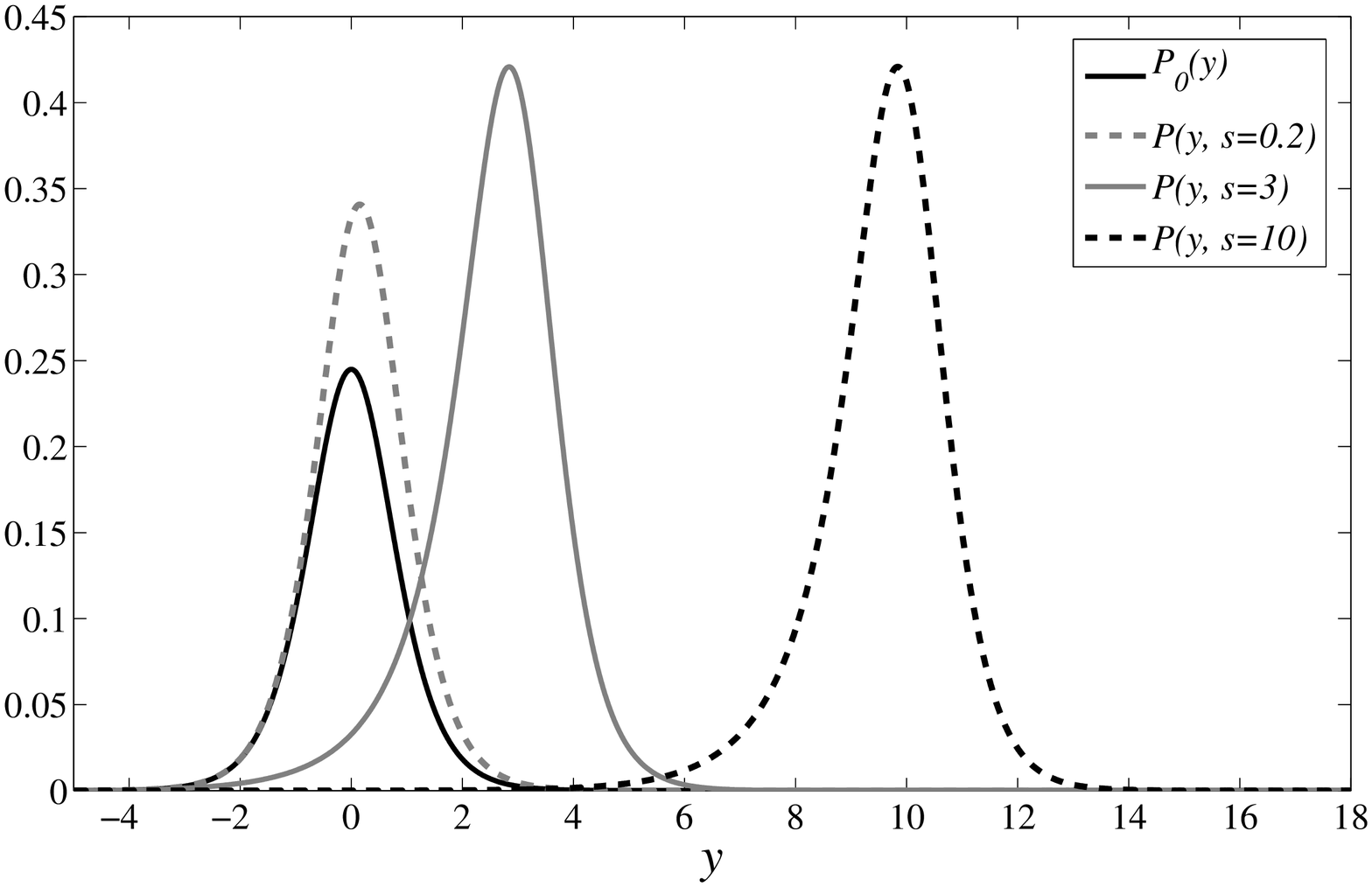}
\end{center}
\caption{Spatial dispersion $P(y, s)$ of the agents with velocity $V_+=+1$ for different times $s=[0; 0.2; 3; 10]$. The initial conditions are given by $P_{0}(y)=Q_{0}(y)=\frac{1}{8\gamma}\left[\mathrm{tanh}(y+\gamma)-\mathrm{tanh}(y-\gamma)\right]$, $\alpha=0$, $\beta=2$ and $\gamma=\frac{1}{4}$.} \label{spatial_dispersion}
\end{figure}
Fig. \ref{spatial_dispersion}. The joint evolutions of  $P(y, s)$ and $Q(y, s)$ (representing agents with velocity $V_+=+1$ and $V_-=-1$ respectively) are drawn in 
\begin{figure}[h]
\begin{center}
\includegraphics[width=14 cm,height=8.5 cm]{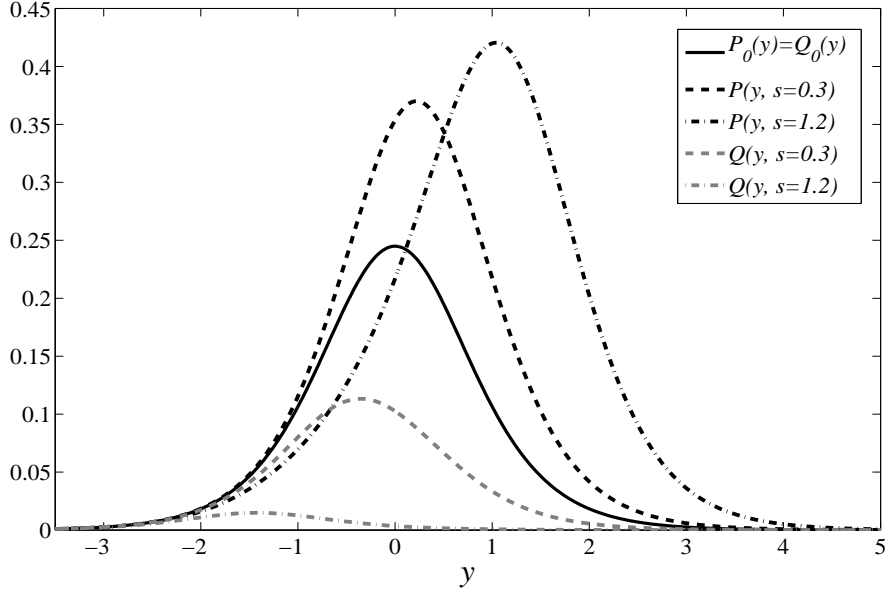}
\end{center}
\caption{Spatial dispersions $P(y, s)$ and $Q(y, s)$ of the agents with velocity $V_+=+1$ and $V_-=-1$ respectively, for times $s=[0; 0.3; 1.2]$. The initial conditions are given by $P_{0}(y)=Q_{0}(y)=\frac{1}{8\gamma}\left[\mathrm{tanh}(y+\gamma)-\mathrm{tanh}(y-\gamma)\right]$, $\alpha=0$, $\beta=2$ and $\gamma=\frac{1}{4}$.} \label{spatial_dispersion_with_q_interaction}
\end{figure}
Fig. \ref{spatial_dispersion_with_q_interaction}.\\
\\
In the present configuration, the two types of agents have an identical initial spatial distribution ({\it i.e.} $P_{0}(y)=Q_{0}(y)$, $\forall y$). Hence, half of the agents have initially the velocity $V_+=+1$, the other half having the velocity $V_-=-1$ ({\it i.e.} $\int_{\mathbb{R}}  P_{0}(y)dy=\int_{\mathbb{R}}  Q_{0}(y)dy=\frac{1}{2}$). The spatio-temporal behavior of the solution given by Eqs.\eqref{interaction_1} and \eqref{interaction_2} can be split into two different time phases. For short times of the dynamics, the overlap  between $P(y,s)$ and $Q(y,s)$ is non-null  ({\it i.e.} the two populations of agents ${\cal A}_+$ and ${\cal A}_-$ are not spatially segregated), thus leading to strong  imitation processes between the agents.   Agents are changing their velocity from $V_-=-1$ to $V_+=+1$ due to both imitation and  spontaneous transitions. Accordingly, $Q(y,s)$ is gradually depopulated  for the benefit of  $P(y,s)$. The imitation processes are decreasing as the overlap between  $P(y,s)$ and $Q(y,s)$ gets smaller, but  the rate $\beta$ of spontaneous transitions from $V_-=-1$ to $V_+=+1$ remains itself constant with time. In the time asymptotic regime, almost  all the agents have adopted velocity $V_+=+1$, leading $P(y,s)$ to behave as  a probability density uniformly traveling  with velocity $V_+=+1$ towards the positive  $\mathbb{R}$-axis.\\
\\ 
In 
\begin{figure}[h]
\begin{center}
\includegraphics[width=11.4 cm,height=7.5 cm]{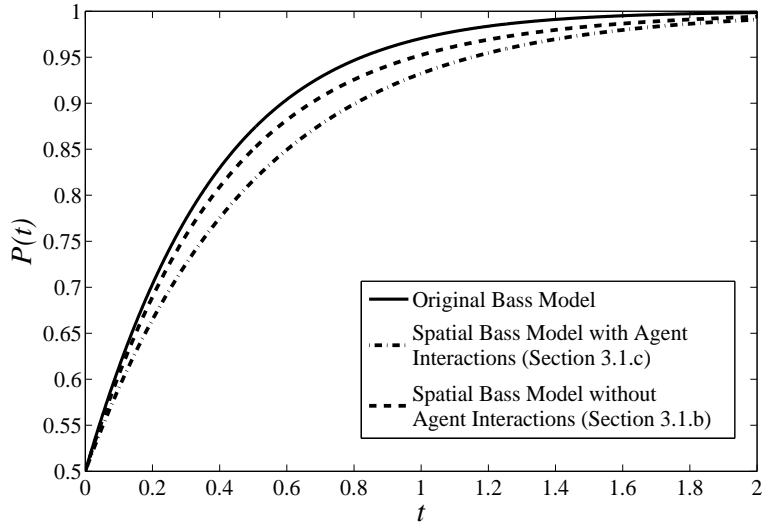}
\end{center}
\caption{Temporal evolution of the overall adoption rate $P(t)$ {\it i)} for the original Bass' model, {\it ii)} for our spatial Bass' model when there are imitation processes (Section 3.1.c) and {\it iii)} for our spatial Bass' model when there are no imitation processes because the two populations of agents ${\cal A}_+$ and ${\cal A}_-$ are immediately segregated (Section 3.1.b).} \label{bass_comparison}
\end{figure}
Fig. \ref{bass_comparison}, the temporal evolution of the overall adoption rate $P(t)= \int_{\mathbb{R}}  P(y,t) dy$ of the agents is illustrated and compared to the one observed for the Bass' original model. For the present configuration, the overall adoption rate stands between $\beta$, the rate observed for our spatial Bass' model in absence of imitation processes (Section 3.1.b), and $\beta + 1$, the rate obtained for the original aggregated Bass' model. Hence, in general, the overall adoption rate of our spatial Bass' model will be equal to $\beta +\epsilon(t)$, $\epsilon(t) \in ]0, 1]$, with $\epsilon(t)$  depending on the initial distributions of the agents and on the model parameters. Remember that the imitation rate (and hence the overall adoption rate) is controlled by the number of {\it neighbors} that each agent  effectively observes during the imitation process. In the aggregated (original) Bass' model, this number is maximum as each agent systematically observes the global population of agents. In the limiting regime of Section 3.1.b, the number of observed agents is equal to 0 as the two populations of agents ${\cal A}_+$ and ${\cal A}_-$ are initially and hence permanently segregated, thus allowing no imitation processes.  

\end{itemize}
\section{Conclusion and Perspectives}

\noindent Often, agents' remoteness may naturally reduce the efficiency of  imitation processes, a feature  that is totally absent in the original aggregated Bass' approach. Incorporating a spatial dimension into the Bass' dynamics is however not a minor extension. It transforms indeed a nonlinear single dimension dynamics into a nonlinear infinite dimensional field dynamics for which, in general, no solution methods are available. It is hence remarkable that our spatial generalization of  the original Bass' dynamics  leads to a class of models for which the evolution of  the resulting spatio-temporal patterns can be  exactly  calculated.  Indeed, the quadratic  nonlinearity,  due here to the underlying imitation mechanism, coincides with the collision term found in  a solvable Boltzmann equation, the RW dynamics,  used for gas models in mathematical physics. Our paper points out that the  RW  dynamics, solvable via  a linearizing  logarithmic transformation,  offers a unique, synthetic and exact  modeling framework to study nonlinear features generated by spatio-temporal imitation processes in economics systems.\\
\\
\noindent  In this paper,  we mainly  focused on Bass' dynamics which does not allow back transitions ({\it i.e.} $\alpha =0$, implying no transitions from $V_+$ to $V_-$). Allowing $\alpha > 0$, regimes involving shock waves with propagating velocities $w$ emerge and the velocity range is ${\beta - \alpha  \over \beta +\alpha  } < w <1$,  \cite{ruijgrok}.  When a dominant spontaneous tendency to stay in the $V_-$ state exists ({\it i.e.} $\alpha > \beta$), imitations enhance the spontaneous  $\beta$-flow from  $V_- \mapsto V_+$  and may lead to a time-independent  ({\it i.e.} $w=0$), shock type inhomogeneous solution. Depending on the  initial conditions, this marginal stationary $w=0$ solution separates two regimes of shocks  propagating with either positive or negative velocities, a rich dynamical behavior that deserves further investigations in economy.

\end{document}